\documentclass[12pt]{iopart}

\usepackage{iopams}
\usepackage{graphicx}
\usepackage{txfonts}

\begin{document}

\title[]{Implicit and electrostatic Particle-in-cell/Monte Carlo model
in two-dimensional and axisymmetric geometry II: Self-bias voltage
effects in capacitively coupled plasmas}

\author{Wei Jiang$^1$, Hong-yu Wang$^{1,2}$, Zhen-hua Bi$^1$ and You-nian Wang$^1$}

\address{$^1$School of Physics and Optoelectronic Technology, Dalian
University of Technology, Dalian, 116024, P. R. China}
\address{$^2$Depart of Physics, Anshan Normal University, Anshan,
 114007, P. R. China}
\ead{\mailto{ynwang@dlut.edu.cn}}

\begin{abstract}
With an implicit Particle-in-cell/Monte Carlo model, capacitively
coupled plasmas are studied in two-dimensional and axisymmetric
geometry. Self-bias dc voltage effects are self-consistently
considered. Due to finite length effects,the self-bias dc voltages
show sophisticating relations with the electrode areas.
Two-dimensional kinetic effects are also illuminated. Compare to the
fluid mode, PIC/MC model is numerical-diffusion-free and thus finer
properties of the plasmas are simulated.
\end{abstract}
\submitto{\PSST}  \vspace{2pc}
\maketitle

\section{Introduction}
Capacitively coupled plasmas (CCP) processing is the mainstream
technology for etching and deposition devices in semiconductor
industry \cite{Lieberman05,Makabe06,Kushner09}. Besides their
applications in semiconductor industry, many physical processes
involved in CCP, are still not fully understood and therefore
attracted many researchers, both from the academy and the industry.
Except many analytical models to understand the physics in CCP
qualitatively, there are two ways \cite{Kim05,Dijk09} to study the
plasma process in the reactors quantitatively: fluid/Monte Carlo
(MC) hybrid method and Particle-in-cell/Monte Carlo (PIC/MC) method.

PIC/MC model is widely adopted in academy research because it has
fewer assumptions. However, PIC/MC model is very computationally
expensive. As a result, up to now, most PIC/MC simulations for CCP
were only done in 1D geometry
\cite{Georgieva03,Boyle04,Kim04,Kawamura06,Bronold07, Donko09,
Jiang09}. There are only several open reports about standard 2D
simulations. Vahedi \cite{Vahedi93b} presents the first 2D results
based on direct implicit PIC/MC model, but in planar (X-Y) geometry.
Recently, Kawamura \cite{Kawamura08} studied the dc/rf discharges
with Vahedi's model in the same geometry. First 2D axisymmetric
analysis was given by Nanbu \cite{Wakayama03}, for which the code is
executed on their supercomputers. Recently we also conducted 2D
axisymmetric simulations for CCP\cite{Wang09}, but in a very small
zone. Although 1D PIC/MC simulations can reveal most the physics in
CCP, such as plasma density, sheath thickness and heating rate,
however, some characteristics of CCP is inherently two dimensional.
For example, magnetized CCP \cite{Lee07,Kim07,Leray09} and very high
frequency CCP \cite{Chabert05,Lee08}. Of course, the most general
dimensional effect is the self-bias dc voltage \cite{Yonemura05,
Bultinck08}.

Due to more electrode surfaces are naturally grounded than driven,
most CCPs are asymmetric. Because of the exitance of the blocking
capacitor, negative self-bias dc voltage will build up on the rf
powered electrode. Self-bias dc voltage is mainly determinate by the
geometric factors, namely, the ratio of powered-to-grounded
electrode area $A_{a}/A_{b}$, where $a$ donates the rf electrode and
$b$ donates the grounded electrode. Lieberman
\cite{Lieberman89,Lieberman90} first proposed an analytic spherical
shell model, and obtained good agreements with the experimental
results. This model is expanded to low frequency case by Kawamura
\cite{Kawamura99}. Boswell \cite{Smith97} first adopted 1D PIC/MC
spherical simulations and investigated the evolution of the bias
voltage. With similar model, Yonemura \cite{Yonemura04} studied the
self bias voltage systematically, and the simulation results are
well consistent with measurements and Lieberman's theory. This
method is also adopted by many other researcher. Besides, self-bias
dc voltage effects in CCP had also been well included and studied by
fluid model \cite{Rauf97,Rauf99}. However, for 2D geometries, the
voltage ratio does not simply scale as a power of the area ratio,
but depends in a complicated way on many other effects. In addition,
finite plasma length and nonlocal effects \cite{Kaganovich02} have
been shown to play essential roles in understanding the low pressure
plasmas, where the electron mean free length is comparable to even
larger than the electrode spacing. Electron heating and energy
dissipating mechanisms may be significantly modified. Nevertheless,
theoretical and numerical investigations of such effects are only
carried out in 1D, one can anticipate finite radius may introduce
similar effects. Therefore more 2D PIC/MC simulations are highly
desired to give more insights into this problem, especially for the
kinetics and non-local effects.

This is the second one of our two serial papers. In the first one of
our two serial papers, we have developed an implicit and
electrostatic PIC/MC model in two-dimensional and axisymmetric
geometry. In this paper, we studied the self-bias dc voltage depends
on the radius of the rf powered electrode with this model. We will
present the physical and numerical parameters in Sec.2. Simulation
results are given and compared with some fluid results in Sec 3.
Finally, discussions and a brief summary are presented in Sec.4.

\section{Computational parameters}
\begin{figure}
\includegraphics[scale=0.7]{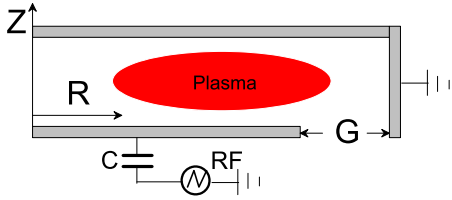}
\caption{\label{Device} Schematic of the simulation for CCP.}
\end{figure}

The schematic of the simulation for CCP is shown in Fig.
\ref{Device}. The physics parameters of the benchmark problem are
similar to our benchmark problems, excepted that the external
circuit is included. The frequency of rf source $\omega_{rf}$ is
$13.56$MHz. Voltage source is applied to the electrode at $z=0 cm$
with waveform of $V_{rf}=200 \sin \omega_{rf}t$, through a blocking
capacitor of $C=300$pF. Since the capacitor is large, the discharge
is essentially voltage driven \cite{Kawamura99}. Argon gas is used
with the pressure of $100$mTorr and temperature of $300K$. We
consider elastic, excitation and ionization collisions for electrons
and elastic and charge transfer collisions for Ar+ ions,
respectively. The electrodes spacing is $L=2$cm, the radius of the
outer cylinder is $R=8$cm and the gap between the lower power
electrode to the grounded outer cylinder is $G1=1$,$G2=2$,$G4=4$cm,
respectively.

Square cells are used, thus Z direction is uniformly divided to 64
cells and 256 cells are in R. The space and time steps are fixed to
all simulations, $\Delta x=0.02/64 m$, $\Delta t_e=1\times10^{-10}$s
and $\Delta t_i=10\Delta t_e$. Note here we did not subcycle the MC
process to avoid violation the condition of the null collision
method for ions. We adopted somewhat a little larger time step to
save time, which would result somewhat lower density by a factor of
about $0.6\sim0.8$, but most the physics, such as the sheath
thickness, is still preserved. The initial density is uniform of
$5\times10^{15}m^{-3}$ for all cells and 200 particles are placed
randomly within one cell. During the simulations, totally about
$5-7\times10^6$ particles are traced. One simulation will take about
40 to 60 hours for 1000 rf periods before convergence, in 4 nodes of
our cluster. All results are given by averaging over one rf period
after reaching equilibrium.

The numerical schemes are also chosen similar to the benchmark
problems, except that the electrons are subcycled. The self-bias dc
voltages are calculated self consistently from PIC/MC model with
Vahedi \cite{Vahedi97} model. For comparison, fluid model
simulations, which have be detailed discussed in our former
paper\cite{Bi09}, are performed with identical physics parameters,
except that the voltage waveforms from the PIC/MC model are used. A
typical fluid simulation will take only about 18 hours on a single
Intel E2160 CPU, or about $1/20\sim1/30$ computation cost of
implicit PIC/MC simulations.

\section{Simulation results}
\subsection{Self-bias dc voltage}
\begin{figure}
\includegraphics[scale=0.7]{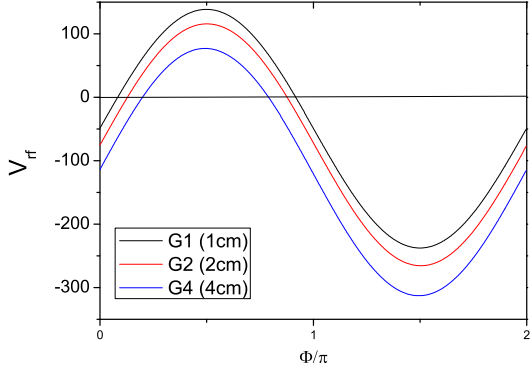}
\caption{\label{Voltage} Voltage waveforms on the rf powered
electrode.}
\end{figure}

The calculated voltages on the rf powered electrode are plotted in
Fig.\ref{Voltage}, it is very clear that the voltages are in
$V=V_{0}\sin \omega_{rf}t+V_{dc}$ form, where $V_{0}$ is 188V, 191V
and 195V, $V_{dc}$ is -49V, -74V and -117V for G1, G2 and G4 case,
respectively. Due to the potential drop on the capacitor, $V_{0}$ is
slightly smaller than $V_{rf}$, and $V_{dc}$ increases with
decreasing rf electrode radius. No higher harmonic oscillations are
observed in our simulations.

\begin{figure}
\includegraphics[scale=0.7]{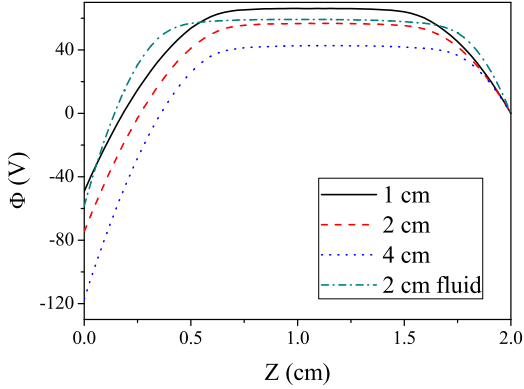}
\caption{\label{Phi_Z} Axial cross-sectional profiles for $\Phi$ at
$R=2cm$ for different gap lengths from PIC/MC model. We also plotted
the result from fluid model with the gap length of $2cm$ for
comparison.}
\end{figure}

For clarity, cross-sectional profiles of $\Phi$ at $R=2cm$ for
different gap lengths are shown in Fig.\ref{Phi_Z}. The potential
drop near the rf electrode $V_a$ and the potential drop near the
grounded electrode $V_b$, can be readily read from the figure. Here
we have the electrode surface area ratio $A_b/A_a=1.3$, $1.8$ and
$4$, for the G1,G2 and G4 case, respectively. The $V_a/V_b$ and
$A_b/A_a$ has the relation
\begin{equation}
\frac{V_a}{V_b}=(\frac{A_b}{A_a})^q.
\end{equation}

\begin{figure}
\includegraphics[scale=0.7]{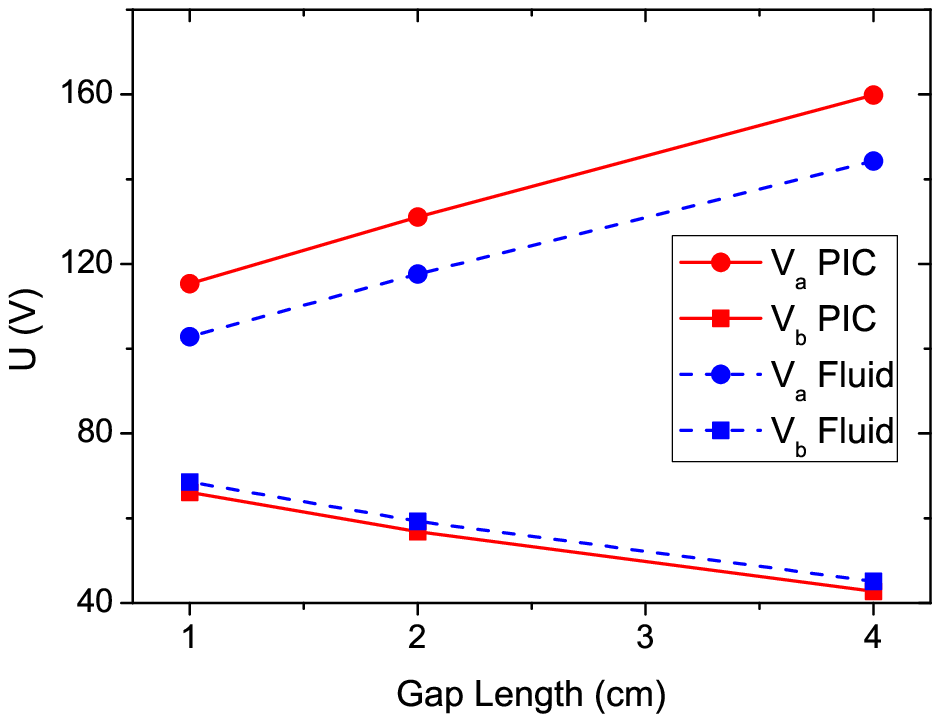}
\includegraphics[scale=0.7]{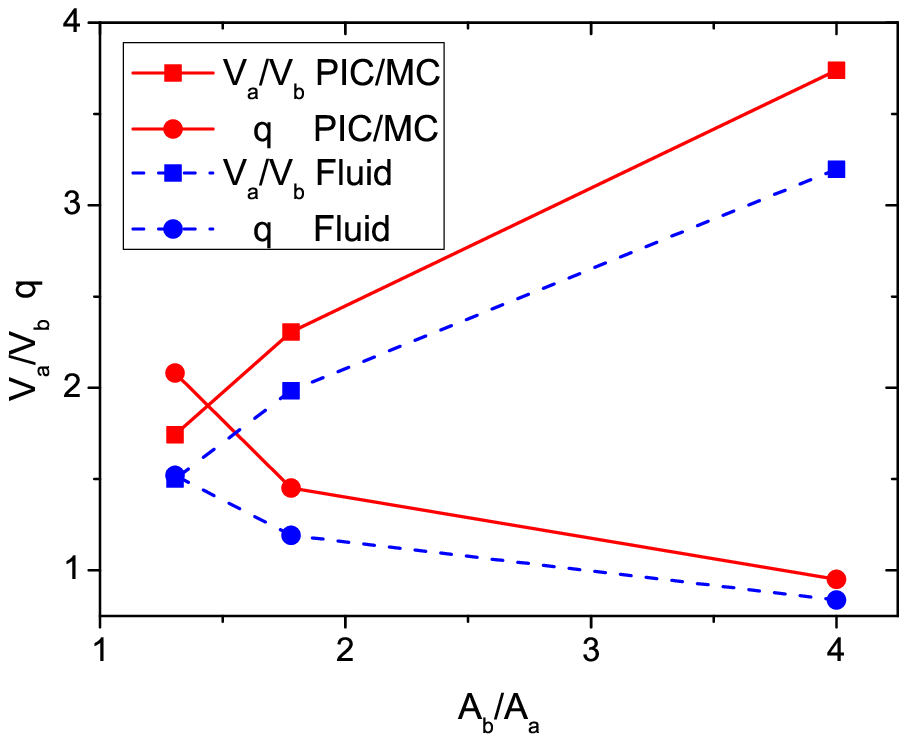}
\caption{\label{Q} (a)Average potential drop near the rf electrode
$V_a$ and potential drop near the grounded electrode $V_b$; and (b)
$V_a/V_b$ and power $q$ as a function of ${A_b}/{A_a}$. The solid
line is from PIC/MC model and the dashed line is from fluid model}
\end{figure}

We plotted $V_a$, $V_b$ and $q$ as function of gap lengths in
Fig.\ref{Q}. It can be concluded that $V_a$ increases and $V_b$
decreases with increasing ${A_b}/{A_a}$. The $V_a$ from the fluid
model is smaller than that from PIC/MC model, while the $V_b$ is
larger, even the fluid model simulation adopted voltage from PIC/MC
results. The $q$ is also smaller from fluid model.

The key issue to understand the problem is the electron and ion mean
free length, $\lambda_e\approx3 cm$ and $\lambda_i\approx0.03 cm$.
For $G=1cm$ case (${A_b}/{A_a}=1.31$), the $q=2.08$, very close to
the collisional case where $q=2.21$ \cite{Lieberman89}. In this
case, the gap is narrow, the side wall sheath is thin thus has
little effects on the bulk plasmas, the 1D spherical shell model
will give a good estimation for $q$. For $G=2cm$ case
(${A_b}/{A_a}=1.77$), the $q=1.45$, same to value $q=1.45$ given by
Lieberman's finite radius model\cite{Lieberman90}. In this case, the
gap length is moderate,as well as the side wall sheath length. For
$G=4cm$ case (${A_b}/{A_a}=4$), the electrode radius is comparable
to the electron mean free length, local effect will be significant
and the plasma is heavily disturbed by the side sheath, $q$ is can
not be predicted by the analytical model, we have $q=0.95$. We can
conclude that finite radius effects are always tend to decrease $q$.

\begin{figure}
\includegraphics[scale=0.4]{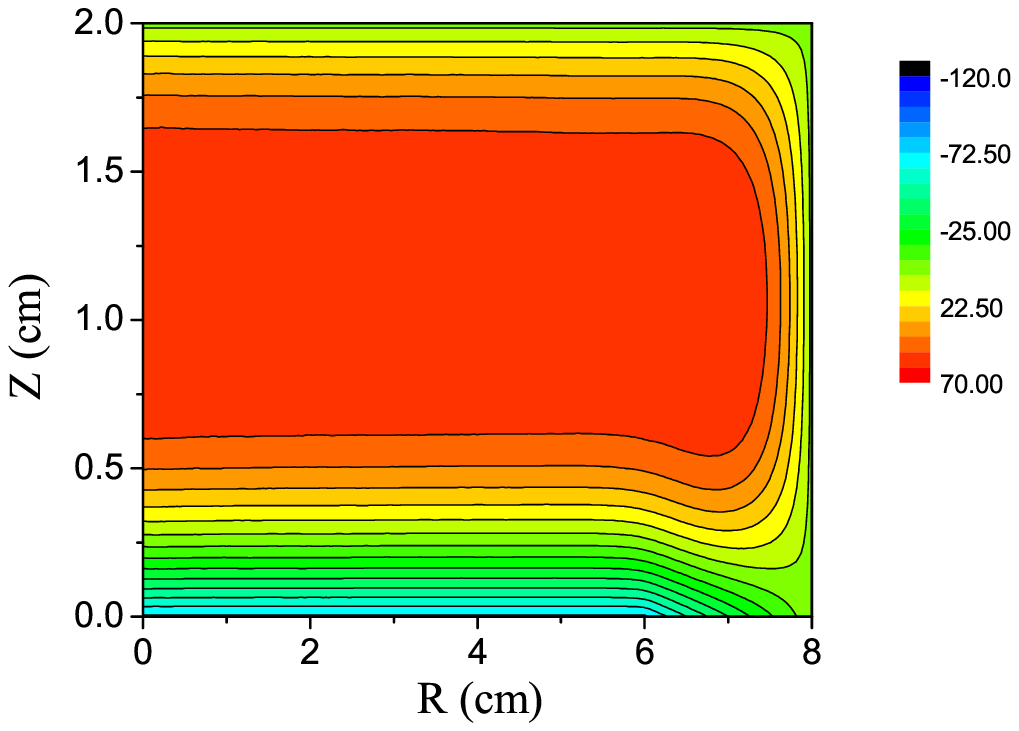}
\includegraphics[scale=0.4]{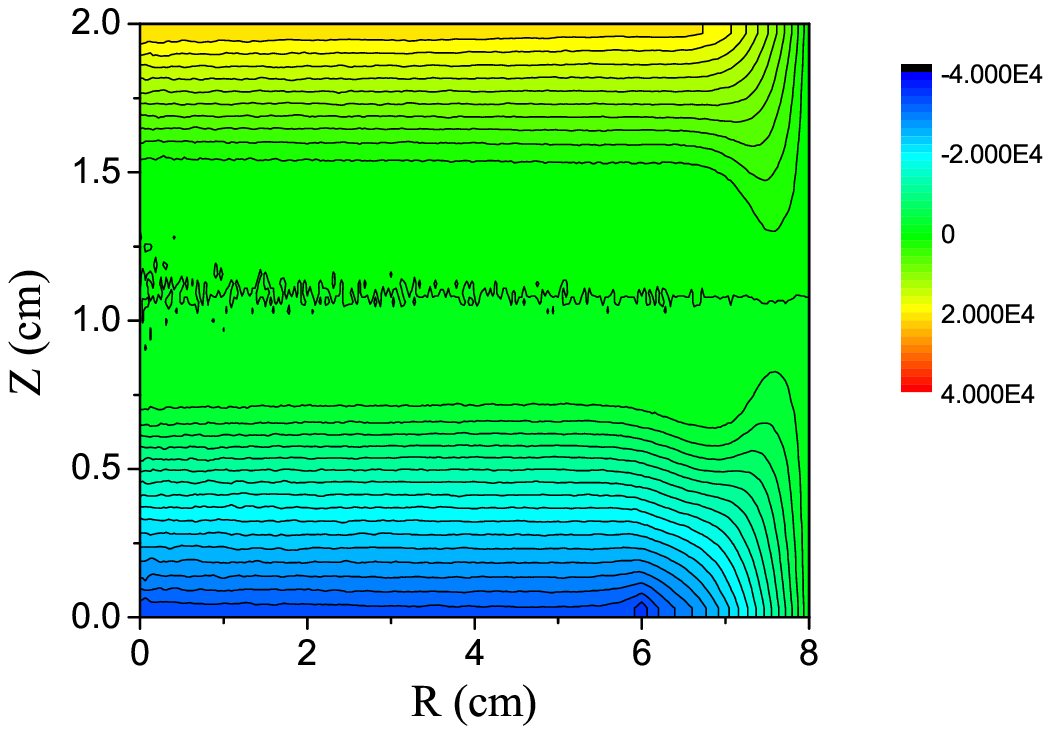}
\includegraphics[scale=0.4]{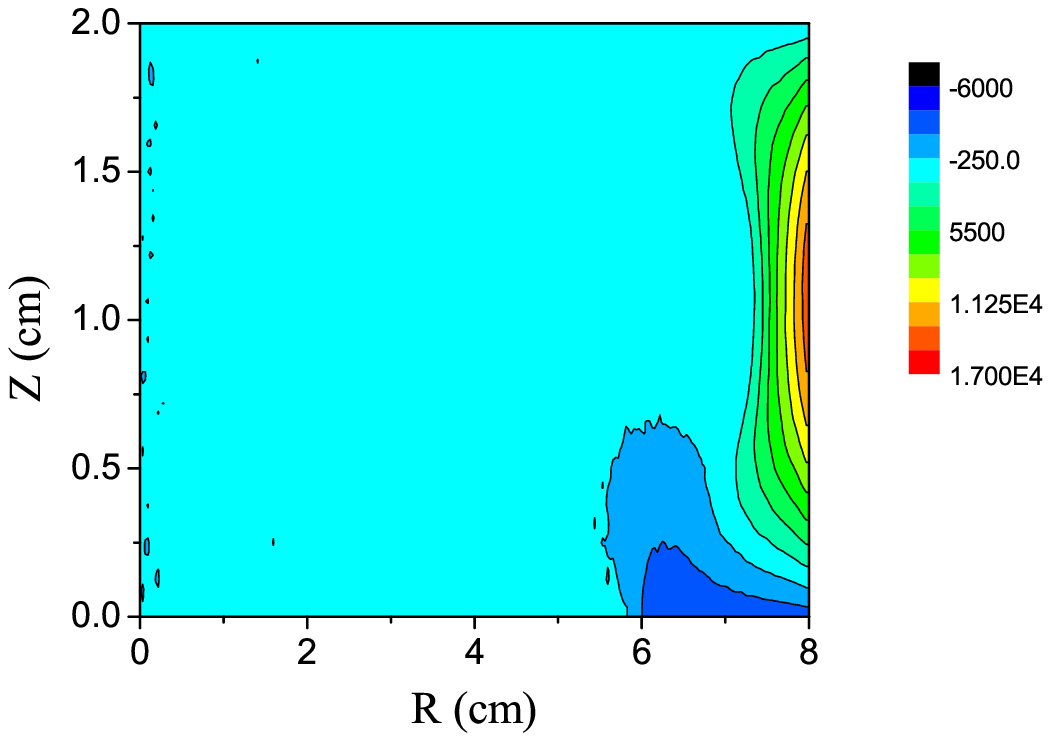}
\caption{\label{Field_PIC} 2D average (a)$\Phi$(V), (b)$E_z$ (V/m)
and (c)$E_r$ (V/m) profiles from PIC/MC model. The gap length is
$2cm$.}
\end{figure}

\begin{figure}
\includegraphics[scale=0.4]{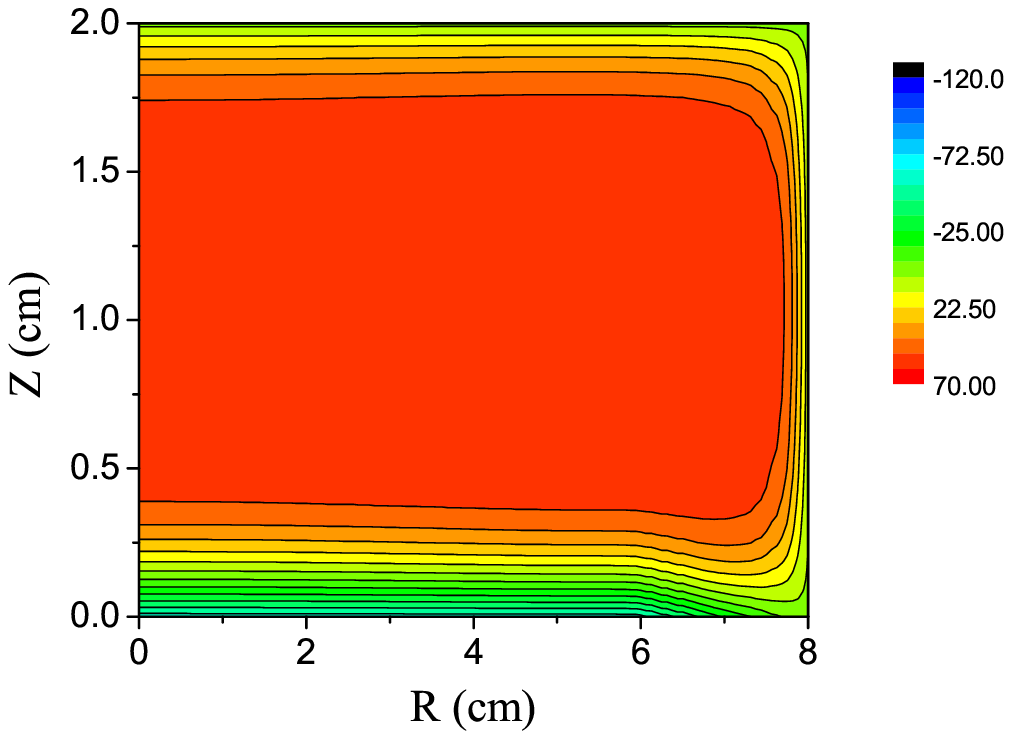}
\includegraphics[scale=0.4]{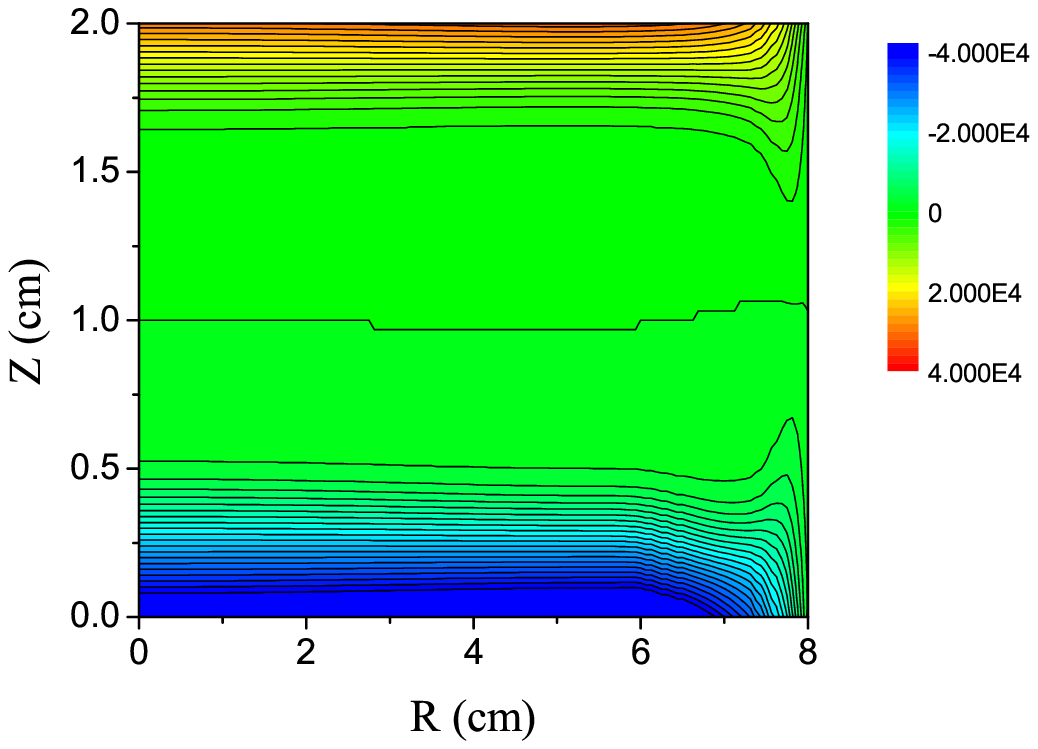}
\includegraphics[scale=0.4]{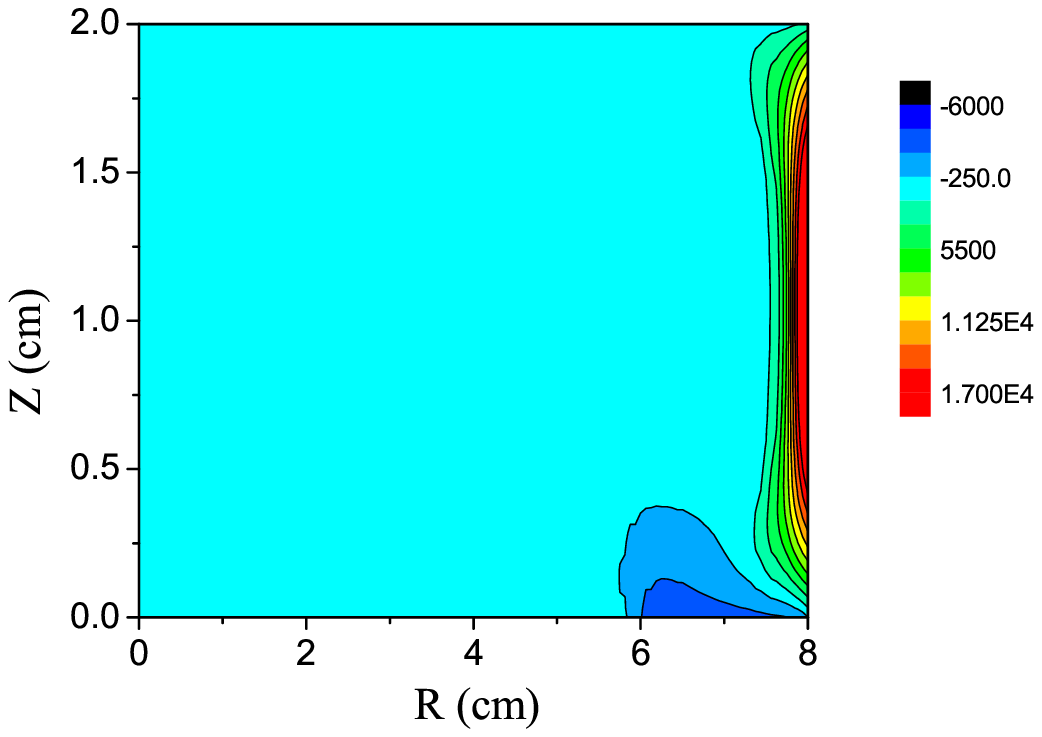}
\caption{\label{Field_Fluid} 2D average (a)$\Phi$(V), (b)$E_z$ (V/m)
and (c)$E_r$ (V/m) profiles from fluid model. The gap length is
$2cm$.}
\end{figure}

The period average potential and field $\Phi$, $E_z$ and $E_r$ from
PIC/MC model are depicted in Fig.\ref{Field_PIC}, corresponding
results from fluid model are shown in Fig.\ref{Field_Fluid}, both
for G2 case.  all for G2 case. Unlike the 1-D symmetric simulation,
the potential $\Phi$ is more negative since there are self-bias
voltages. The $E_z$ in the axial direction has a structure similar
to the 1-D simulation, except the drop near the side wall. The $E_r$
is small, except near the side wall. Due to the ion sheath near the
wall, the $E_r$ is very large and positive there, except that $E_r$
is negative near the gap since the self-bias voltage is negative.
Fluid model gives very similar results. But the fluid results have
larger plateau in the center. And the $E_z$ given by PIC/MC model is
slightly smaller than $E_z$ given by fluid model.


\subsection{Plasma density}
\begin{figure}
\includegraphics[scale=0.6]{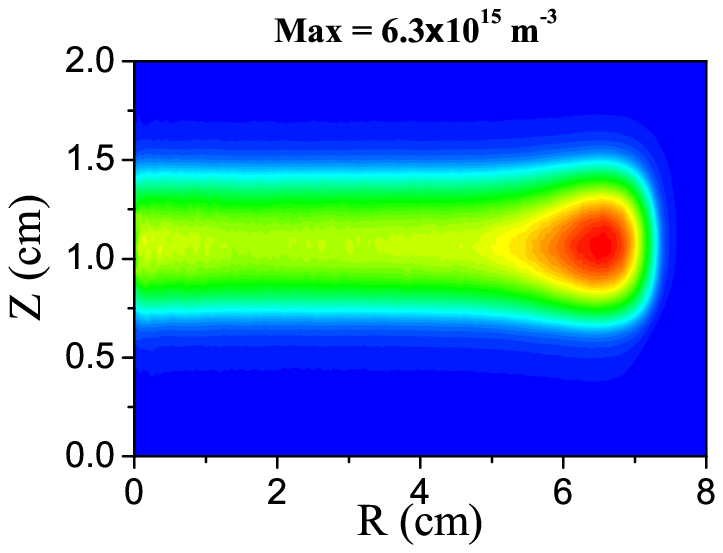}
\includegraphics[scale=0.6]{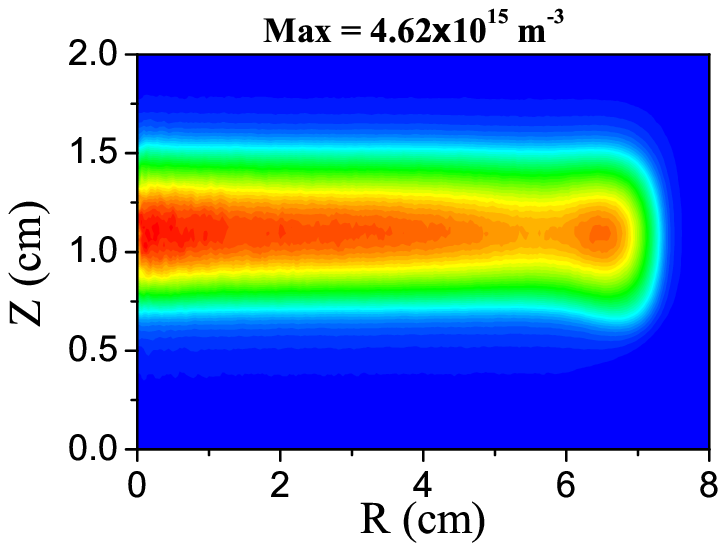}
\includegraphics[scale=0.6]{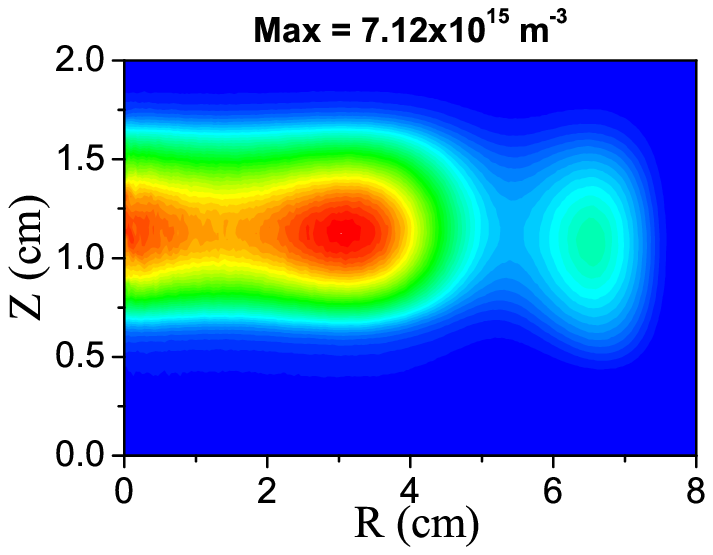}
\caption{\label{Ne_PIC} Average electron density profiles from
PIC/MC model. The gap lengths are (a)$1cm$, (b)$2cm$ and (c)$4cm$.}
\end{figure}

\begin{figure}
\includegraphics[scale=0.6]{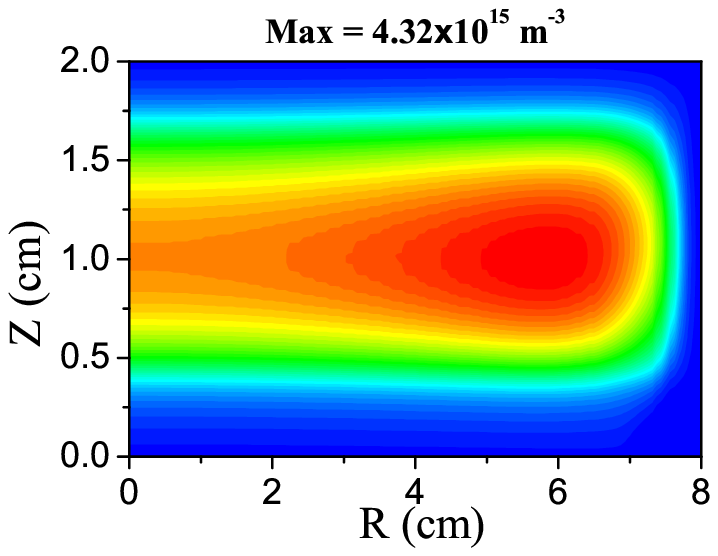}
\includegraphics[scale=0.6]{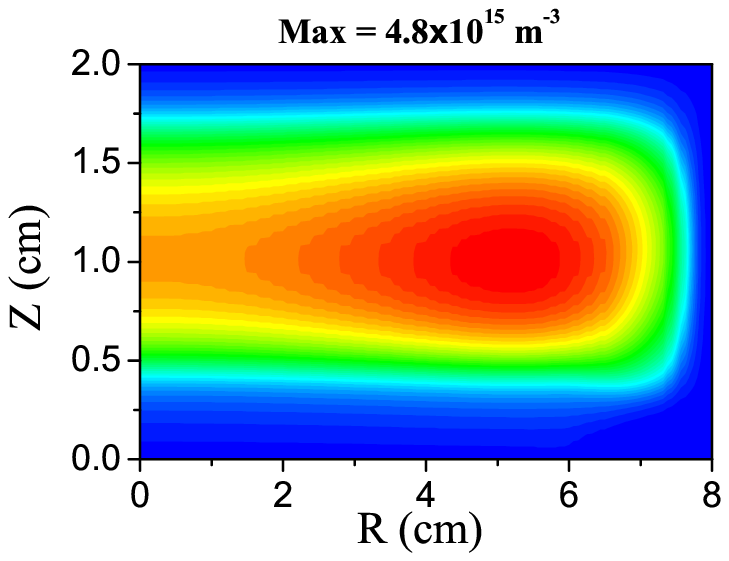}
\includegraphics[scale=0.6]{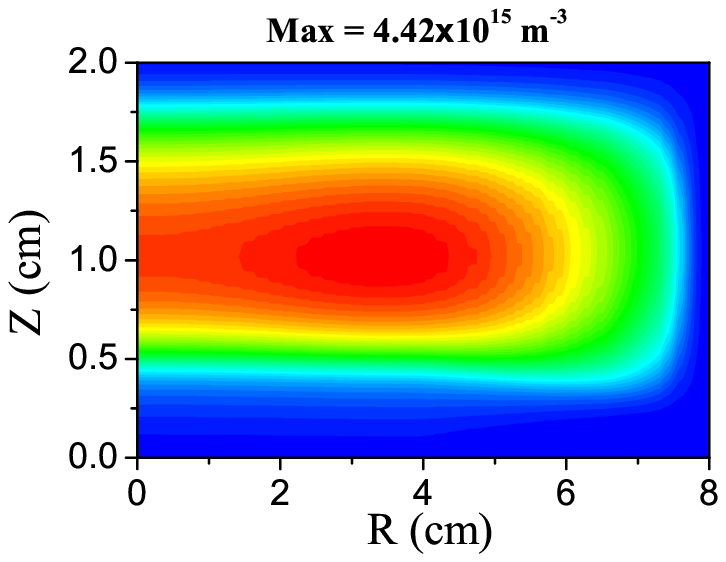}
\caption{\label{Ne_Fluid} Average electron density profiles from
fluid model. The gap lengths are (a)$1cm$, (b)$2cm$ and (c)$4cm$.}
\end{figure}

The electron densities from PIC/MC model are shown in
Fig.\ref{Ne_PIC}, corresponding results from fluid model are shown
in Fig.\ref{Ne_Fluid}, all for different gap length. It can be
clearly seen that both fluid model and PIC/MC model give similar
results for the peak plasma density and the profiles. The axial
cross sections of the density are very similar to the 1D results,
except near the gap.

In fluid results, the density profiles are much more flat and
smooth, there are only one peak near the gap, for all three cases.
This phenomena comes from the well-known numerical diffusion effect.
In Eulerian simulations, the discrete equations always give larger
diffusive coefficients than the original differential equations in
general, even if Flux-Corrected Transport (FCT) method is often
used. As a result, fluid model tends to smooth out all the short
wave length oscillation and lessens the density gradients.

In PIC/MC results, since the kinetics effects are included, the
cases are much more complex. For G1 case, there are only one peak
near the gap, the radial density is nearly constant between the gap
and the axis, like many other fluid simulations have predicted
\cite{Rauf97, Rauf99}. For G2 case, the only difference is the self
bias voltage, compared to the benchmark problem in our first paper.
There are two peaks, the lower one is near the gap, the higher is in
the axis, also similar to the benchmark problems. But in this case,
self bias dc voltage makes the density gradients between the gap and
the axis become smaller. The G4 case has the largest peak density,
and there are three peaks, the most high peak is near the gap, the
lower peak is near the axis, and the lowest is between the gap and
the side wall. As we have discussed, in this case the electron free
length is comparable to the electrode radius, electron can be
bounced by the radial sheath to gain energy (then increase the
densities). On the other hand, when it runs towards the axis, it can
decrease the velocity due to the collision. For the change of the
velocity, it can be accumulated near the axis then two peaks
appears. When the electrons and the ions drift out of the rf powered
electrode, they can be trapped and heated in the side wall sheath,
and thus form the third peak. These peaks, are inherently kinetic
effect, and all are smoothed out in the fluid simulation, even if
fluid model can give reasonable density. It also should be noted
that the maximum density is G4 case, and the the minimum density is
G2 case for PIC/MC model. But the maximum density is G2 case, and
the the minimum density is G4 case for fluid model.

\begin{figure}
\includegraphics[scale=0.7]{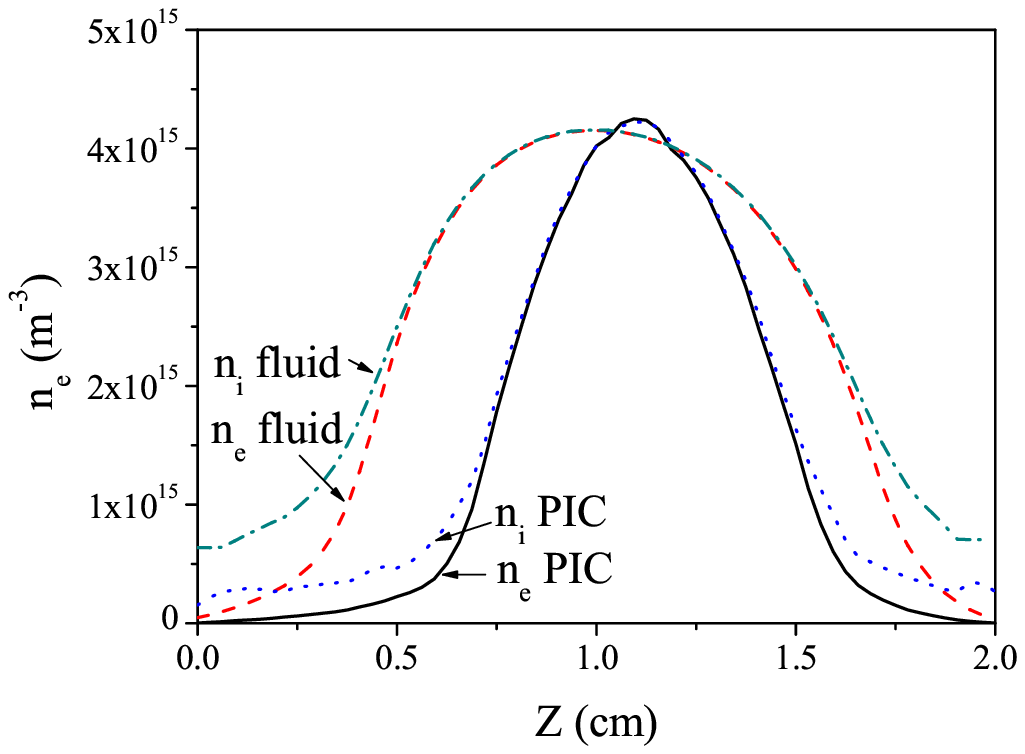}
\includegraphics[scale=0.7]{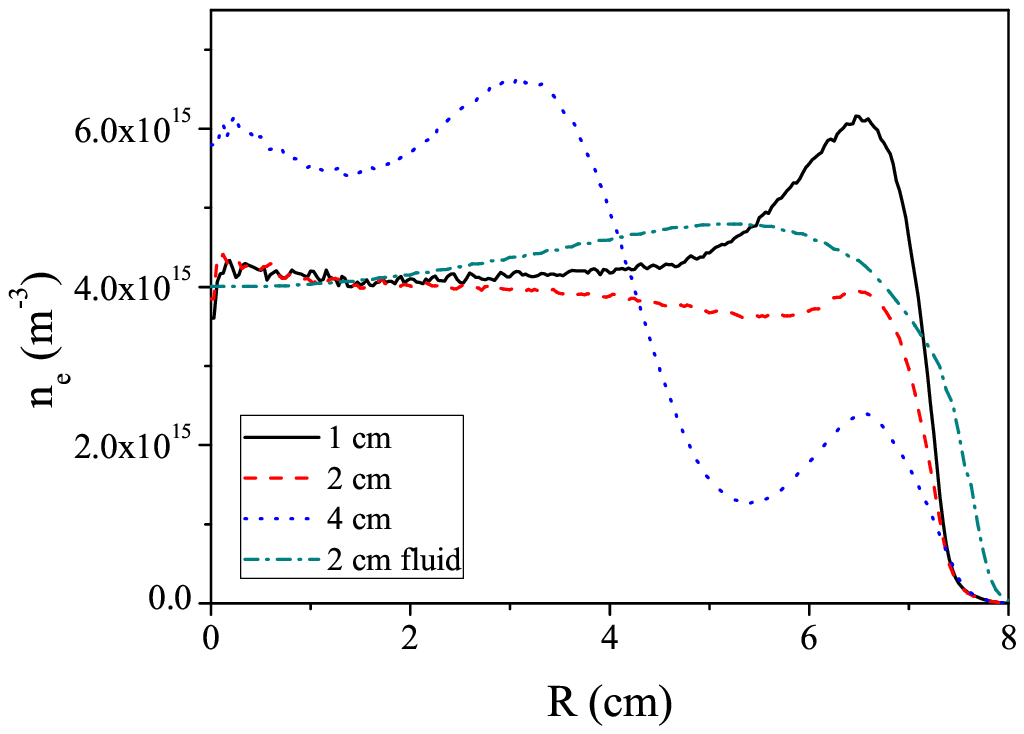}
\caption{\label{Ne_ZR} (a) Axial cross-sectional profile at $R=2cm$
of electron and ion density and for $G=2cm$; (b) radial
cross-sectional profile of electron density at $Z=1cm$ for different
gap lengths, both from PIC/MC model. We also plotted the same
results from fluid model with $G=2cm$ for comparison}
\end{figure}

For clarity, we plotted the axial and radial cross-sectional
profiles for the electron and ion density for different gap lengths
Fig.\ref{Ne_ZR}. For axial profile, PIC/MC model give steeper
results, as well as the sheath thickness, again, the reason is the
numerical diffusion in the fluid model. Due to the dc bias voltage,
the peak is more close to the grounded electrode, but the profile is
still close to the 1D results. Note here because we have adopted
large time steps in PIC/MC simulations, we have underestimate the
max density by a factor of about $0.6\sim0.8$.

For radial profile, all three cases give similar side sheath
thickness, larger than that from the fluid model. The reason is the
same. G4 case gives the largest density. As we have discussed,
enhanced heating due to the finite radial length, is responsible for
the result. G2 case has similar density to the G1 case and the fluid
results in the axis, but gives smaller value at larger radius. It
seems that, in this case, finite radial length effect tend to
decrease the density, by increasing the side wall ion loss. It
should be noted that the peaks near the wall all appeared at the
same radial position.

\subsection{Electron temperature}
\begin{figure}
\includegraphics[scale=0.7]{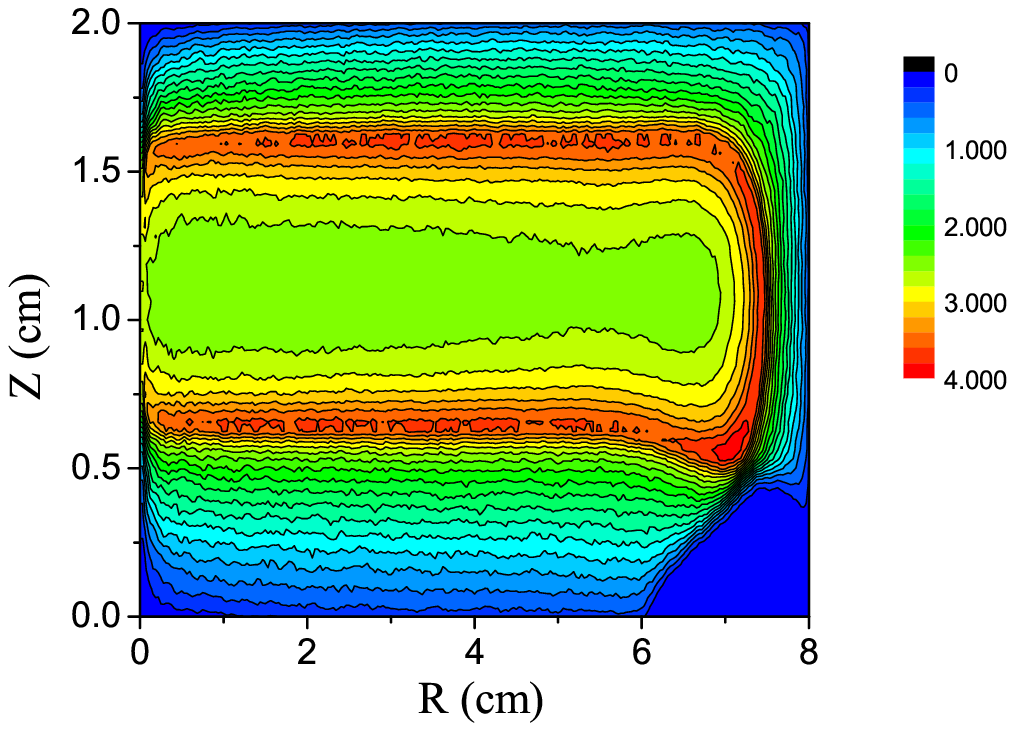}
\includegraphics[scale=0.7]{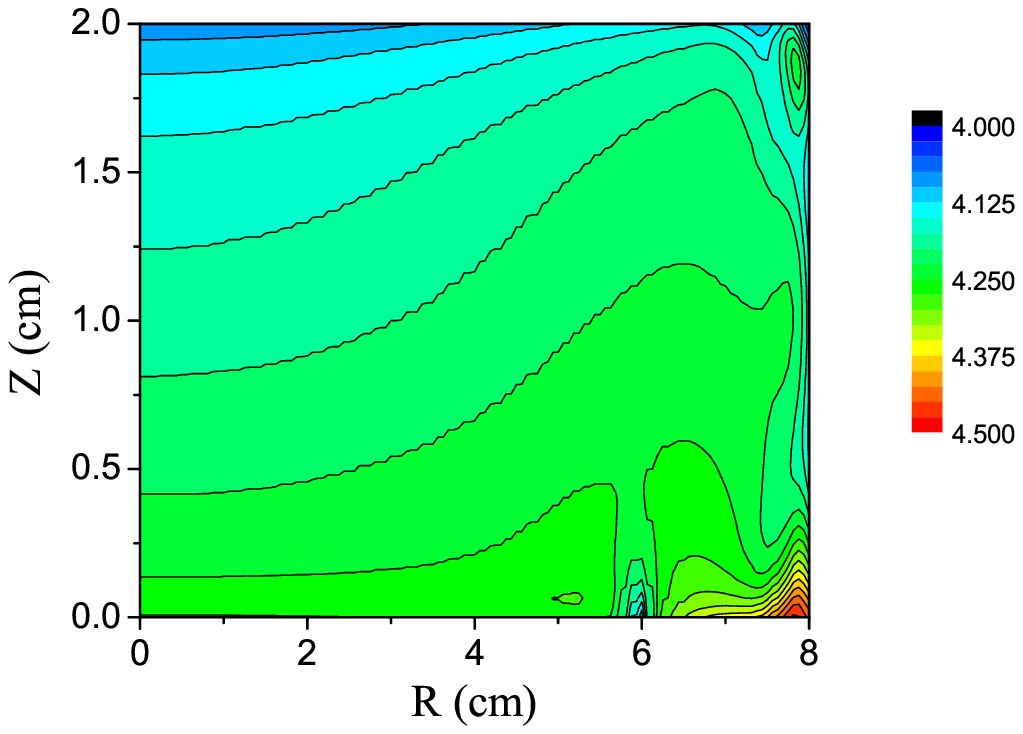}
\caption{\label{Te_2cm} 2D average electron temperature profiles
from PIC/MC model(a) and fluid model (b). The gap length is all
$2cm$.}
\end{figure}

We depicted the 2D average electron temperature profiles from PIC/MC
model and fluid model in Fig.\ref{Te_2cm}, for gap length of $2cm$.
The differences are significant. For fluid model, the density are
about $4eV$ and nearly constant over very large area. But for PIC/MC
model, the profile is saddle like in axial direction. This implies
the electron heating mostly occurs in the sheath. Note there is also
a peak in the side wall sheath, implying electron heated there, by
the oscillating side wall sheath and the $E_r$. The most significant
difference is near the gap corner, fluid model predict a electron
temperature enhancement there, while PIC/MC model give zero
temperature because there is a small zone without electrons in all
time. Again, this is a result of numerical diffusion effect of fluid
model.

For clarity we also plotted the axial cross-sectional profiles at
$R=2cm$ of electron temperature for different gap lengths. Again,
the profiles are very similar to 1D results, a saddle like form. The
temperature given by the fluid model is nearly constant and larger
than those given by PIC/MC model. At larger self dc voltage, the
peak will tend to move towards the grounded electrode.

\begin{figure}
\includegraphics[scale=0.7]{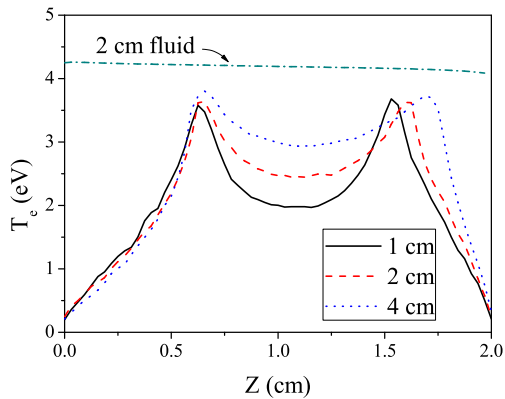}
\caption{\label{Te_Z} Axial cross-sectional profiles at $R=2cm$ of
electron temperature for different gap lengths, both from PIC/MC
model. We also plotted the same results from fluid model with
$G=2cm$ for comparison}
\end{figure}

\subsection{Ion flux and energy distributions}
\begin{figure}
\includegraphics[scale=0.7]{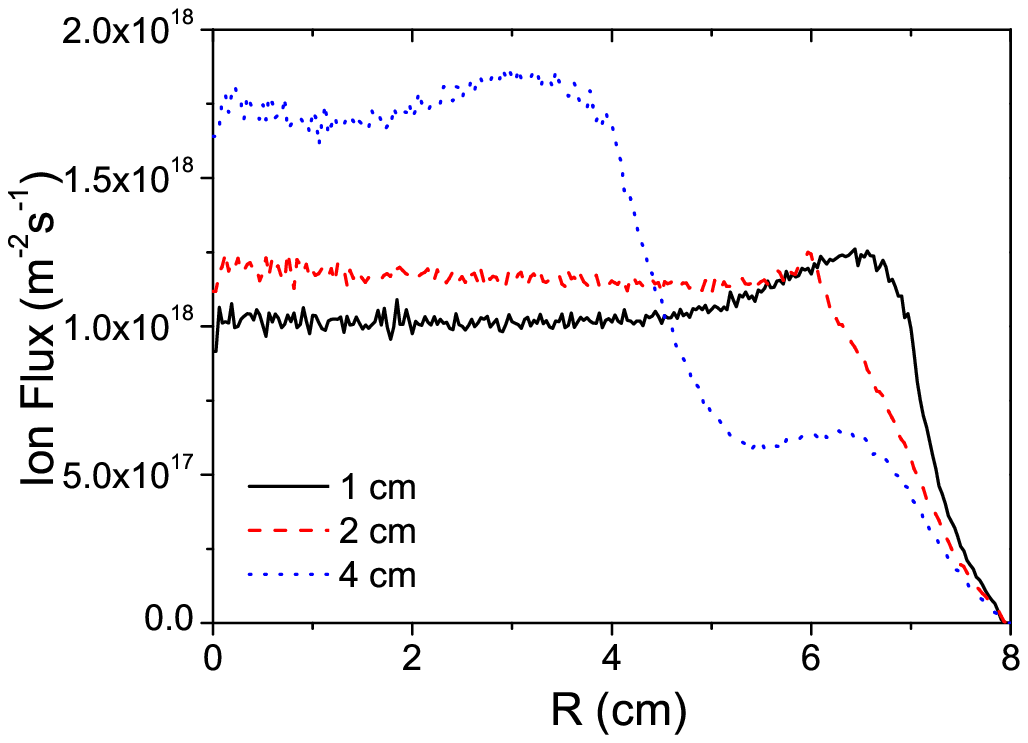}
\includegraphics[scale=0.7]{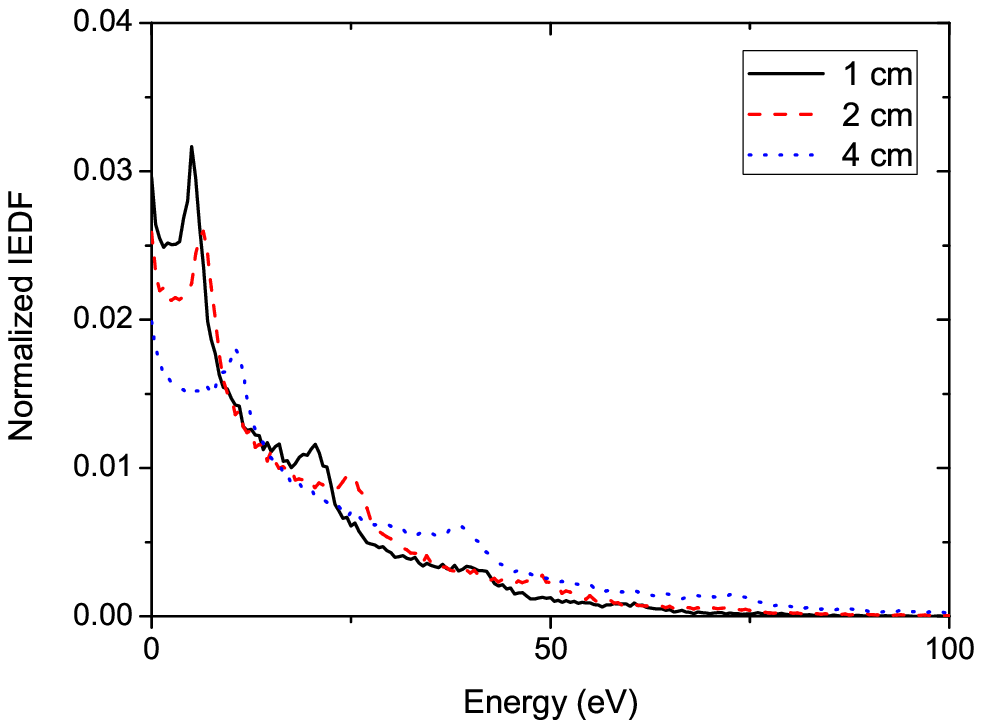}
\caption{\label{Flux} (a)Radial distribution of ion flux onto the rf
powered electrode; and (b) ion energy distribution functions on the
rf powered electrode, both from PIC/MC model.}
\end{figure}

Fig.\ref{Flux}(a) shows the radial distribution of ion flux onto the
rf powered electrode for three gap lengths. As can be seen, although
G4 case gives the largest flux, the flux is not very uniform, while
for the G1 and G2 case, the flux is uniform over the rf powered
electrode. It should be noted that the ion flux is larger for G2
case, although the density is smaller. It seems the larger side wall
sheath tend to increase the flux to the electrode.

Fig.\ref{Flux}(b) shows the ion energy distribution functions
(IEDFs) on the rf powered electrode. Due to the pressure is high,
the two peaks in the IEDFs are not very clear; and larger gaps give
larger average ion energy, because the self bias dc voltage is
larger, as we have discussed.

\section{Discussions and summary}
In this paper, we have studied the self-bias dc voltage with 2D
PIC/MC model. At small gap length, the dc voltage can be well
estimated by 1D spherical shell model\cite{Lieberman89}; at moderate
gap length, the dc voltage can be well estimated from the infinite
radius model\cite{Lieberman90}. However, at small gap length, the dc
voltage can not be estimated by the analytic mode, due to electron
nonlocal behavior will dominate.

Due to the numerical diffusion effect, although it can give
reasonable density values and profiles, fluid model tends to smooth
out all the short wave length oscillation and lessen the density
gradients. The density and electron temperature profiles given by
the PIC/MC model are more steep. Due to nonlocal and kinetic
effects, there are several peaks in the density profiles.
The simulations validate both PIC/MC model we
have adopted and the fluid model.

However, PIC/MC model still has many shortcomings compare to fluid
model, which may severely constrain the applications of this model.
For example, PIC/MC model is computationally expensive, and is very
hard or even practically impossible to couple with chemical reaction
model and neutral gas model\cite{Takekida06}. Nevertheless, through
PIC/MC model, exactly plasma behavior can be predicted, kinetics
effects can be preserved. We are trying to give more insights into
the physics of CCP with this model.

\section*{Acknowledgments}
This work was supported by the National Natural Science Foundation
of China (No.10635010).

\end{document}